\documentclass[PRB,noshowpacs,amsmath,amssymb,superscriptaddress,twocolumn]{revtex4}
\usepackage[T1]{fontenc}
\usepackage[latin1]{inputenc}
\usepackage{amsmath}
\usepackage{amssymb}
\usepackage{graphicx}
\usepackage{epsfig}
\usepackage{bm}
\usepackage{dsfont}
\usepackage{color}

\newcommand{\bea}{\begin{eqnarray}}
\newcommand{\eea}{\end{eqnarray}}
\newcommand{\be}{\begin{equation}}
\newcommand{\ee}{\end{equation}}
\newcommand{\bc}{\begin{center}}
\newcommand{\ec}{\end{center}}

\newcommand{\bwt}{\begin{widetext}}  
\newcommand{\ewt}{\end{widetext}}

\begin{document}

\title{Time-resolved collapse and revival of the Kondo state near a quantum phase transition}

\author{Christoph Wetli}
\affiliation{Department of Materials, ETH Zurich, 8093 Zurich, Switzerland}

\author{Shovon Pal}
\affiliation{Department of Materials, ETH Zurich, 8093 Zurich, Switzerland}

\author{Johann Kroha}
\affiliation{Physikalisches Institut and Bethe Center for Theoretical Physics,
Universit\"at Bonn, Nussallee 12, D-53115 Bonn, Germany} 
\affiliation{Center for Correlated Matter, Zhejiang University, 
Hangzhou, Zhejiang 310058, China} 

\author{Kristin Kliemt}
\affiliation{Physikalisches Institut, Goethe-Universit\"{a}t Frankfurt, Max-von-Laue-Str.\ 1, 60438 Frankfurt, Germany}

\author{Cornelius Krellner}
\affiliation{Physikalisches Institut, Goethe-Universit\"{a}t Frankfurt, Max-von-Laue-Str.\ 1, 60438 Frankfurt, Germany}

\author{Oliver Stockert}
\affiliation{Max Planck Institute for Chemical Physics of Solids, 01187 Dresden, Germany}

\author{Hilbert von L\"ohneysen}
\affiliation{Institut f\"ur Festk\"orperphysik and Physikalisches Institut, Karlsruhe Institute of Technology, 76021 Karlsruhe, Germany}

\author{Manfred Fiebig}
\affiliation{Department of Materials, ETH Zurich, 8093 Zurich, Switzerland}


\begin{abstract}
\end{abstract}

\maketitle

{\bf 
One of the most successful paradigms of many-body physics is the concept of
quasiparticles: excitations in strongly interacting matter behaving like
weakly interacting particles in free space. Quasiparticles in metals are very
robust objects. Nevertheless, when a system's ground state undergoes a
qualitative change at a quantum critical point (QCP) \cite{Loehneysen07}, the
quasiparticles may disintegrate and give way to an exotic quantum-fluid state
of matter. The nature of this breakdown is intensely debated
\cite{Si01,Coleman01,Senthil04,Woelfle11}, because the emergent quantum fluid
dominates material properties up to high temperatures and might even be
related to the occurrence of superconductivity in some compounds \cite{Stockert11}. Here we trace the dynamics of heavy-fermion quasiparticles in CeCu$_{6-x}$Au$_{x}$ and monitor their evolution towards the QCP in time-resolved experiments, supported by many-body calculations. A terahertz pulse disrupts the many-body heavy-fermion state. Under emission of a delayed, phase-coherent terahertz reflex the heavy-fermion state recovers, with a coherence time 100 times longer than typically associated with correlated metals \cite{Wolf98,Kummer12}. The quasiparticle weight collapses towards the QCP, yet its formation temperature remains constant -- phenomena believed to be mutually exclusive. Coexistence in the same experiment calls for revisions in our view on quantum criticality.
}

All across condensed-matter physics, coherent excitations of many-body systems can be described in a simple picture of strongly renormalized, but weakly interacting particle-like objects -- the so-called quasiparticles. In metals, for example, heavy Landau fermionic quasiparticles of remarkable stability are enforced by the Pauli principle. Despite the Pauli stabilization, however, these heavy-fermion quasiparticles are amenable to disintegration near a quantum phase transition (QPT) \cite{Woelfle18} because of their low binding energy, thus opening an avenue to novel quantum states of matter, governed by quantum fluctuations. At a QPT, near zero temperature, the critical fluctuations are generated by the energy allowed by Heisenberg's uncertainty principle rather than thermal excitations. As a consequence, non-Fermi-liquid behaviour, frustrated magnetism or unconventional superconductivity may emerge around the associated QCP.

\begin{figure}[b]
\centering
\vspace*{-0.3cm}
\includegraphics[width=\linewidth]{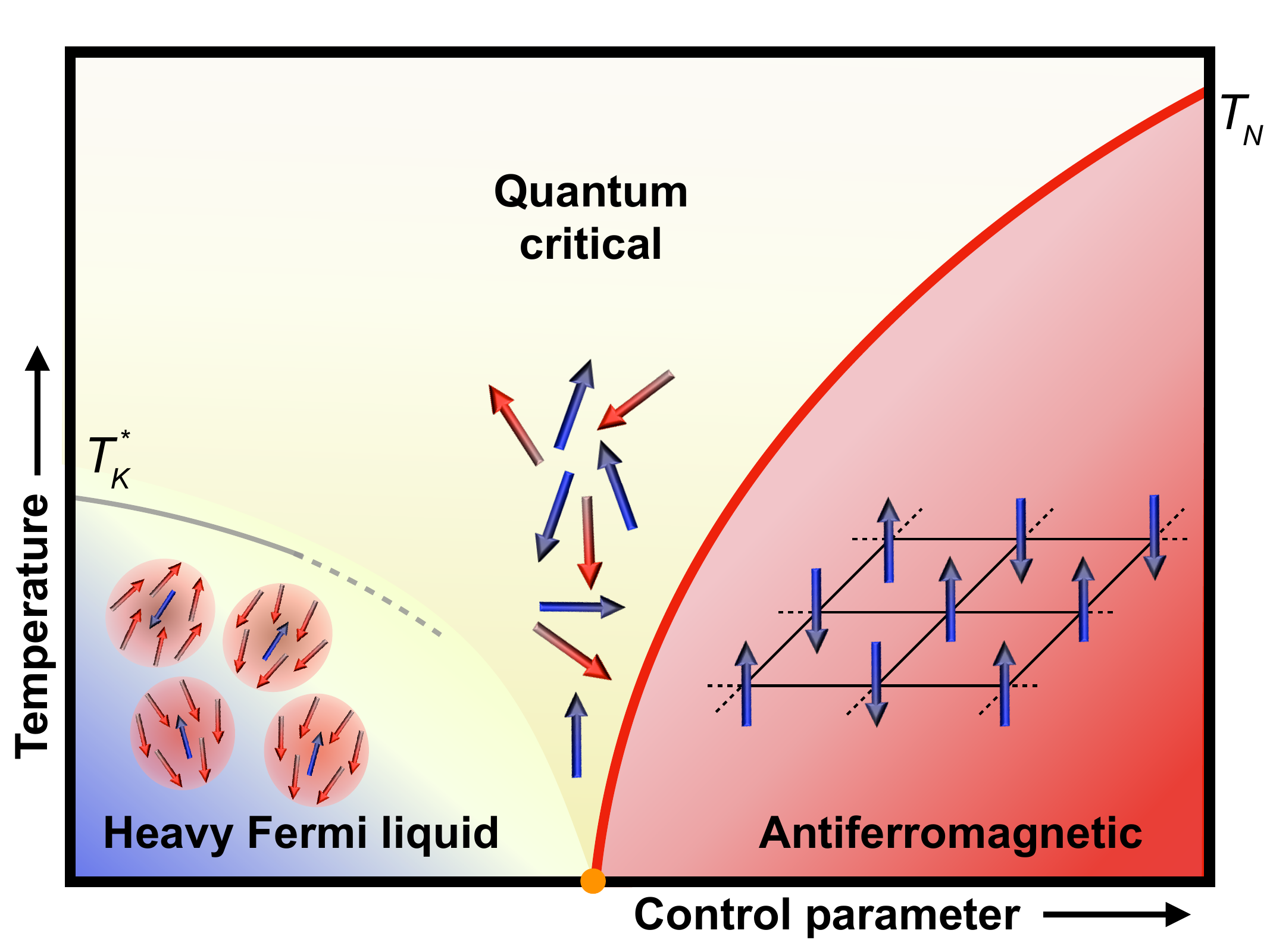}
\caption{\label{phdiag} \textbf{Heavy-fermion quasiparticles near a QCP.} We consider a scenario as in CeCu$_{6-x}$Au$_{x}$, where the quasiparticles disintegrate near the QCP, giving way to an exotic state of matter dominated by quantum fluctuations. In the quantum-critical region (yellow) the quantum energy of fluctuations is larger than the thermal energy $k_BT$. As a consequence,
the system can counter-intuitively enter the quantum-critical region when the temperature is raised.}
\end{figure}

In heavy-fermion compounds, the $4f$ magnetic moments localized on rare-earth ions in the crystal lattice are exchange-coupled to the electron spins residing in the conduction band \cite{Loehneysen07}. Towards low temperatures, the conduction electrons form singlets with the $4f$ spins. The energy scale for this singlet formation defines the Kondo-lattice temperature $T_K^*$, a crossover temperature that typically is of the order of 10~K. Thus, instead of establishing long-range magnetic order mediated by the Ruderman-Kittel-Kasuya-Yosida (RKKY) interaction \cite{Ruderman54,Kasuya56,Yosida57}, a paramagnetic Fermi liquid phase is formed. This Kondo state \cite{Kondo64,Hewson93} is characterized by a sharp resonance in the $4f$ electron spectrum at the Fermi energy. Typically slightly below $T_K^*$, these resonances become lattice-coherent and form a narrow energy band of  
\onecolumngrid

\begin{figure}[t]
\centering
\includegraphics[width=0.8\textwidth]{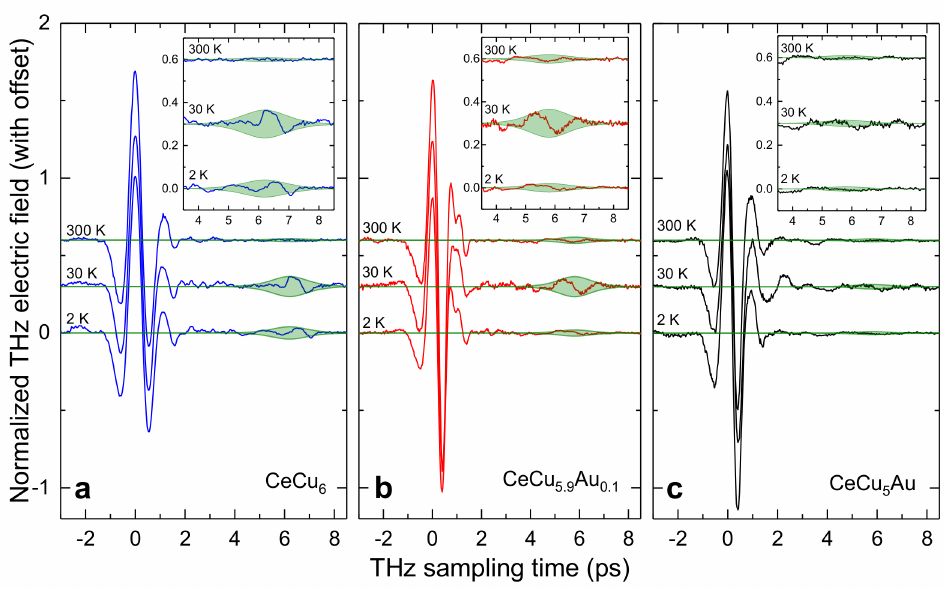}
\caption{\label{thz} \textbf{Time-resolved terahertz reflectivity of the
    heavy-fermion system CeCu$_{6-x}$Au$_x$.} textbf{a--c}, 
  Time dependence of the
  normalized terahertz electric-field amplitude at three different temperatures
  (data vertically displaced) for reflection from the heavy-Fermi-liquid
  compound CeCu$_6$ ({\bf a}), the quantum-critical CeCu$_{5.9}$Au$_{0.1}$
  ({\bf b}) and the antiferromagnetic compound CeCu$_{5}$Au ({\bf c}). Each
  time trace is normalized such that its entire integrated intensity is unity
  (see Methods). The echo pulse is highlighted in the insets. The green-shaded
  area shows the envelope of this time-delayed terahertz reflex which contains
  coherent and certain background contributions as explained in the text. The envelope was derived from a fit of equation~(\ref{eq:envelope}) to the measured data.}
\end{figure}

\twocolumngrid
\noindent long-lived quasiparticles. Its flat dispersion indicates an approximately $10^2$ times heavier effective mass than that of free electrons; hence the name ``heavy fermion''. In the conventional scenario of an antiferromagnetic QCP due to Hertz, Moriya and Millis \cite{Hertz76,Moriya85,Millis93} the heavy quasiparticles remain intact and support critical fluctuations of the magnetization. In many correlated materials, however, this picture is not valid \cite{Schroeder00}. In distinguishing these two options, the behaviour of the Kondo scale $T_K^*$ is an important characteristic as it reveals if and how the heavy quasiparticles disintegrate towards the QCP.

CeCu$_{6-x}$Au$_x$ is one of the best-known heavy-Fermi-liquid compounds
\cite{Loehneysen07,Schroeder00,Stroka93,Ehm07,Klein08}. It is paramagnetic
down to very low temperatures but exhibits short-range intersite magnetic
correlations and hence appears as an ideal system for studying the change in
ground state that a material may undergo near a QCP. Replacing a small
fraction of Cu by Au expands the CeCu$_6$ lattice and leads to a decrease of
the $4f$-conduction electron exchange. This favours the RKKY interaction
\cite{Ruderman54,Kasuya56,Yosida57} between the $4f$ magnetic moments over the
Kondo effect \cite{Doniach77} and induces a QPT to an incommensurate,
antiferromagnetically ordered state \cite{Schroeder94,Loehneysen96} at
$x=0.1$, see Fig.~\ref{phdiag}. A variety of experimental investigations by
thermodynamic and transport measurements
\cite{Marabelli90,Loehneysen06,Loehneysen96}, neutron scattering
\cite{Stroka93,Schroeder00,Schroeder94,Stockert98} and photoemission
spectroscopy \cite{Klein08,Kummer12} revealed valuable aspects of the phases
and properties near the QCP of the CeCu$_{6-x}$Au$_x$ system, but have also
revealed puzzling controversies. In particular, the incompatibility of the
critical behaviour \cite{Loehneysen98,Schroeder00} with the
Hertz-Millis-Moriya scenario \cite{Hertz76,Moriya85,Millis93} points to a
breakdown of the heavy-quasiparticle picture
\cite{Schroeder00,Si01,Coleman01}. In contrast, its preservation
\cite{Hertz76,Moriya85,Millis93,Rosch97} may be concluded from signatures of a
stable, non-vanishing Kondo temperature
\cite{Loehneysen06,Loehneysen98,Klein08}. Such conflicts are caused by the
diversity and, sometimes, indirectness of techniques used to probe QPTs --
techniques that do not permit the measurement of fundamental properties such as the Kondo weight and temperature simultaneously.

In this work, we enter the realm of nonequilibrium dynamics to obtain direct and comprehensive information on the evolution of the quasiparticle picture towards the QCP in CeCu$_{6-x}$Au$_x$. We extinguish part of the Kondo state by irradiation with a terahertz pulse and monitor its time-resolved resurgence. The resulting terahertz echo pulse provides a direct measure for the spectral weight and for the Kondo temperature $T_K^*$ of the heavy-fermion state within a single experiment. We observe that towards the QCP, the coherent spectral weight $w_K^*$ collapses, becoming unobservably small below 5~K. The Kondo energy scale, however, remains constant at $T_K^*\simeq 8$~K. We thus observe a scenario of quasiparticle disintegration where the timescale on which quasiparticles form ($\sim 1/{T_K^*}$) remains finite, whereas the probability for quasiparticle formation ($\sim w_K^*$) collapses. Up to now, these two phenomena were assumed to exclude each other. Now their simultaneous observation calls for revisions in our view on quantum criticality.

We irradiate a set of CeCu$_{6-x}$Au$_x$ samples covering the Fermi-liquid ($x=0$), the quantum-critical ($x=0.1$) and the antiferromagnetic ($x=1$) state with pulses at $0.1-3$~THz. We then record the reflected terahertz wave by free-space electrooptical sampling (see Methods). It is crucial that our terahertz frequencies are just enough to excite electrons from the heavy-fermion band to the conduction band. Our pulses thus disrupt the correlated heavy-fermion state, but they cannot ionize the system, excite electrons out of the tightly bound Ce 4f single-particle states, or induce spin-orbit transitions. The time traces in Figs.~\ref{thz}a-c exhibit an instantaneous reflex ($t=0$) as response of the light conduction electrons. In CeCu$_6$ and CeCu$_{5.9}$Au$_{0.1}$ this is followed by a weaker ``echo'' pulse delayed by $6.2\pm 0.2$~ps and $5.8\pm 0.2$~ps, respectively, while antiferromagnetic \cite{Loehneysen07,Schroeder94,Loehneysen96} CeCu$_5$Au shows a negligible echo pulse. This delayed pulse is not to be confused with trivial etalon (Fabry-Perot) reflexes from the terahertz generation crystal, cryostat windows and so on, which are identified at different delay times (see Methods and Supplementary Information A). Unlike these artefacts, our echo does not appear on a Pt reference sample. It furthermore displays a complex temperature dependence, elaborated on in the following, that none of the trivial reflex pulses show.

\begin{figure}[b]
\centering
\includegraphics[width=\linewidth]{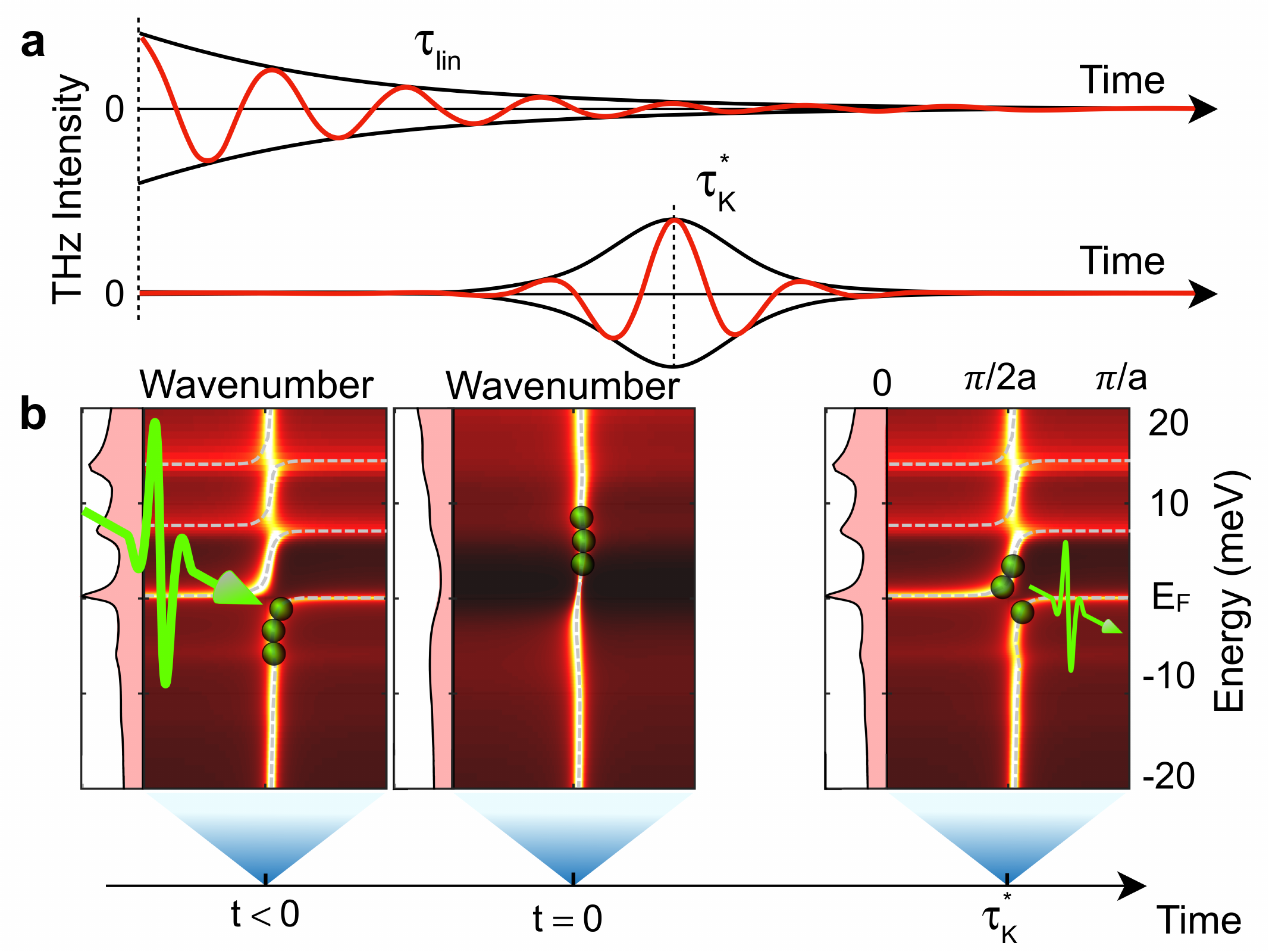}
\caption{\label{relax} \textbf{Dynamics of heavy-fermion quasiparticle
    formation.} {\bf a}, Sketch of the relaxation dynamics after
  photoexcitation of carriers into the conduction band. Top: immediate,
  exponential relaxation of the reemitted electric field resulting from a
  linear rate equation with constant relaxation rate $1/\tau_{lin}$,
  describing, e.g., exciton recombination in semiconductors. Bottom: solution
  (equation~(\ref{eq:envelope})) of the nonlinear rate equation
  (equation~(\ref{eq:rate_eqn})). Here, the heavy-fermion band is of many-body
  origin---that is, its spectral weight and, hence, the relaxation rate dynamically follow the heavy-fermion band population. The pronounced echo pulse with delay time $\tau_K^*$ is a signature of the many-body Kondo physics. {\bf b,} Band structure of a multi-orbital Anderson lattice model, calculated by nonequilibrium dynamical mean field theory (see Methods). Before the terahertz excitation ($t<0$) the nearly flat heavy-fermion band near the Fermi energy $E_F=0$, accompanied by its crystal-field satellites at 8 and 13~meV \cite{Stroka93}, are visible. Excitation with a terahertz pulse ($t=0$) destroys the heavy-fermion state. The associated band collapses and all charge carriers become light electrons. After a time $\tau_{K}^{*}$, resurgence of the heavy-fermion state occurs and the excess energy is released as time-delayed terahertz pulse. The colour scale represents the spectral density in units of the inverse conduction electron band width and ranges linearly from 0.0 (black) to 3.5 (bright yellow). The pink curves on the left of each
panel represent the momentum-integrated, cerium-4f partial density 
of states, ranging from 0.0 to 1.0 in units of the conduction band width.}
\end{figure}

Two features about the echo pulse are striking. First, its oscillatory nature indicates quantum coherent dynamics of the terahertz-excited electrons on the order of 10~ps, about two orders of magnitude longer than coherence times typically associated with optically excited electrons in metals \cite{Wolf98}. Second, the delayed response does not yield instantaneous exponential decay, as one would expect from single-electron relaxation like exciton recombination (Fig.~\ref{relax}a), but a compact pulse, well separated from the instantaneous reflex by a ``dark time''.

The reported \cite{Stroka93,Klein08} CeCu$_6$ Kondo-lattice temperature $T_K^* \approx 6$~K corresponds to a coherence time of $\tau_K^* = h/k_B T_K^* \approx 8$~ps, in remarkable agreement with our observed pulse delay time of about $6$~ps, considering that $T_K^*$ merely marks a crossover scale, not a phase transition temperature. This concordance, the logarithmic temperature dependence of the echo-pulse weight discussed below, and the fact that a delayed pulse is observed only in the heavy-fermion phase but strongly suppressed in the antiferromagnetic phase (see Fig.~\ref{thz}) establish unambiguous evidence for the Kondo-related origin of the delayed pulse: By the terahertz excitation, heavy electrons are excited instantaneously (within the terahertz pulse duration) from the heavy-fermion band into the light part of the conduction band, as shown in Fig.~\ref{relax}b. This transition breaks the Kondo singlet and deletes the associated spectral weight of the heavy band. It then takes the coherence time $\tau_K^*$ to recreate this spectral weight, into which the excited electrons relax with emission of the initially absorbed energy as a delayed terahertz pulse. The oscillatory, time-delayed pulse thus represents the coherent spectral weight of heavy quasiparticles. The phase space for energy- and momentum-conserving scattering of terahertz-excited electrons is drastically reduced compared to optical excitation because of the lower excitation energy. This, together with the large effective heavy-fermion mass, explains the exceptionally long coherence time of $\sim 10$~ps in our experiment.

The unusual echo pulse resulting from this dynamics is quantitatively
described by a semiclassical rate-equation model. According to Fermi's golden
rule, the density of 

\onecolumngrid

\begin{figure}[t]
\centering
\includegraphics[width=0.75\linewidth]{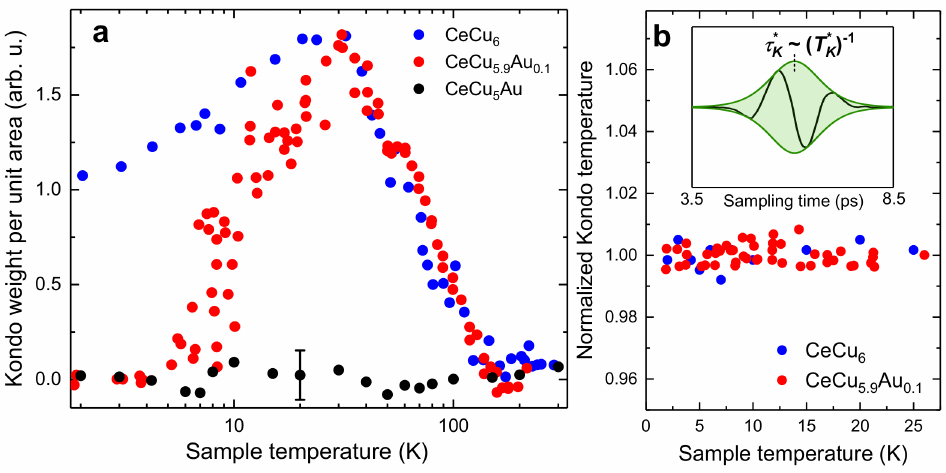}
\caption{\label{qcp} \textbf{Evolution of Kondo weight and Kondo temperature
    towards the QCP in the CeCu$_{6-x}$Au$_x$ system.} {\bf a}, Temperature
  dependence of the Kondo weight per sample area. The weight is derived from
  the integrated intensity of the echo pulse emitted in the $3.5-8.5$~ps
  window (see the insets of Fig.~\ref{thz}) applying the background correction
  described in the text. In the quantum-critical compound
  CeCu$_{5.9}$Au$_{0.1}$ the Kondo weight rises logarithmically between 150
  and 30~K, revealed by the linear slope in the logarithmic plot. Below 30~K
  the Kondo weight collapses, becoming unobservably small at 5~K. In the
  Fermi-liquid compound CeCu$_6$ the behaviour is the same as in
  CeCu$_{5.9}$Au$_{0.1}$ down to 30~K, but the decrease towards lower
  temperature leaves a finite Kondo weight. In antiferromagnetic CeCu$_5$Au,
  the Kondo weight is zero at all temperatures within the statistical error
  (that is, the standard deviation, see error bar). {\bf b}, Evolution of the normalized Kondo temperature towards 0~K for CeCu$_{5.9}$Au$_{0.1}$ and CeCu$_5$Au. Within the statistical error, the Kondo temperature remains constant. The Kondo temperature is derived from the envelope of the echo pulse as sketched in the inset.}
\end{figure}

\twocolumngrid

\noindent 
photoexcited electrons making a transition to the heavy-fermion band per time interval, $d\rho(t)/dt$, is proportional 
to the density of photoexcited electrons $\rho(t)$ and to the spectral density of heavy-fermion final states. The latter is created by the Kondo singlet formation and is therefore proportional to the density of electrons residing in the heavy-fermion band at time $t$, $\rho_0-\rho(t)$, where $\rho_0$ is the density of terahertz-excited electrons from the heavy-fermion band. We thus have a convolution of the recreation of the ground-state spectral density and the repopulation of this recreating spectral density. This leads to a nonlinear rate equation for the normalized density of photoexcited electrons, $n(t)=\rho(t)/\rho_0$,
\begin{eqnarray}\label{eq:rate_eqn}
  \frac{\mathrm{d}n}{\mathrm{d}t}=-\frac{4\pi}{\tau_K^*}(1-n)n\,,
\end{eqnarray}
see Supplementary Information C for a detailed derivation of this simplest possible approach. The electric-field envelope of the emitted terahertz reflex is proportional to this temporal change of occupation, $\overline{E}(t)\sim dn/dt$, and is thus given by the solution of equation~(\ref{eq:rate_eqn}) as,
\begin{eqnarray}\label{eq:envelope}
  \overline{E}(t) =\frac{\overline{E}_0}{\cosh^2[2\pi(t/\tau_K^*-1)]},
\end{eqnarray}
where $\overline{E}_0$ is the maximum pulse amplitude. This envelope is shown in Fig.~\ref{thz}a and agrees well with our experiment. The time-delayed revival of the signal is caused by the dynamical change of spectral weight and is thus an unambiguous signature of correlated many-body dynamics.

We now scrutinize the behaviour of the heavy quasiparticles as the QCP is approached. Our time-resolved terahertz reflectometry allows one to measure both the (integrated) heavy-quasiparticle spectral ``Kondo'' weight $w_K^*$ and the Kondo-lattice temperature $T_K^*$ in one single experiment. Here, $w_K^*$ is obtained from the integrated, background-corrected intensity of the delayed terahertz reflex (see Methods). The Kondo-lattice temperature $T_K^*$ as the energy scale at which the heavy quasiparticles form follows from the inverse pulse delay time as $T_K^* = h/k_B \tau_K^*$. Determination of $T_K^*$ is possible even when the coherent quasiparticle weight vanishes because of incoherent Kondo correlations (reported earlier \cite{Klein08}) that are visible in our time traces as wiggles at the same delay $\tau_K^*$ as the coherent Kondo signature (see discussion in Supplementary Information B).

Figure~\ref{qcp}a shows that the Kondo weight in CeCu$_{5.9}$Au$_{0.1}$ apears below 150~K and rises logarithmically down to about 30~K. Extension of the Kondo effect \cite{Hewson93} to the thermally excited crystal-field states seen in Fig.~\ref{qcp}a \cite{Stroka93,Ehm07} is in line with this high onset temperature. Below 30~K, $w_K^*$ drops continuously, becoming unobservably small near 5~K: direct and striking evidence for quasiparticle disintegration near a QCP. In the non-quantum-critical Fermi-liquid compound CeCu$_6$, we also observe the logarithmic increase of the Kondo weight between 150 and 30~K. However, towards 2~K it drops by only 40\%, reflecting the proximity, yet finite distance to the QCP. In the antiferromagnetic compound CeCu$_5$Au, we have $w_K^*\approx 0$ at all temperatures. Apparently, the coherent Kondo signal is fully suppressed by the antiferromagnetic RKKY interaction \cite{Nejati17} and not only by critical fluctuations near the antiferromagnetic phase transition at 2.2~K \cite{Loehneysen06,Loehneysen07}.

It has been suggested \cite{Si01,Coleman01} that with the quasiparticle breakdown near a QCP the energy scale below which heavy quasiparticles exist vanishes: $T_K^*\to 0$. In particular, observation of a universal quantum-critical fluctuation spectrum evidenced by the so-called ``dynamical scaling'' \cite{Schroeder00}, indicates the absence of an intrinsic energy scale. This was taken as evidence for a vanishing quasiparticle formation scale, even though other experiments \cite{Loehneysen06,Loehneysen98,Klein08} seem to point to a finite $T_K^*$ at the QCP.

Our results, shown in Fig.~\ref{qcp}, reveal that there is no change of the Kondo scale $T_K^*$: Within experimental resolution, the time $\tau_K^*\propto(T_K^*)^{-1}$ it takes the quasiparticles to form remains constant. Nevertheless, the Kondo weight (that is, the probability for the very existence of quasiparticles) collapses at the QCP. Thus, our measurements reconcile the seemingly contradictory phenomena of quasiparticle destruction and their formation energy scale remaining finite. The fact that these two key observations are obtained simultaneously within one single terahertz reflectometry experiment establishes this as a consistent quantum critical scenario. It also appears to be consistent with the observed dynamical scaling in that, although the intrinsic energy scale remains non-zero, its signature at the QCP becomes unobservably weak due to the coherent spectral weight collapse. Although it remains to be seen if this scenario is realized more universally in other heavy-fermion compounds, it suggests a new way of thinking about quantum critical quasiparticle breakdown.

\newpage

\noindent{\bf Acknowledgements}\\[0.1cm]
The authors are grateful for financial support by the SNSF via project No.\ 200021-14708 (M. F., C. W.) and by the DFG via SFB/TR 185 (J. K).

\vspace*{0.4cm}

\noindent{\bf Author Contributions}\\[0.1cm] 
All authors contributed to the discussion and interpretation of the experiment and to the completion of the manuscript. C. W. and S. P. performed the experiment and the data analysis. O. S. and H. v.\ L.  provided the CeCu$_{6-x}$Au$_x$ samples. K. K. and C. K. provided YbRh$_2$Si$_2$ samples for reference experiments. J. K. performed the theoretical analysis. J. K. and M. F. initiated the experiment and supervised the work.

\vspace*{0.4cm}

\noindent{\bf Competing Interests}\\[0.1cm] 
The authors declare no competing interests.

\vspace*{0.4cm}

\noindent{\bf Additional Information}\\[0.1cm]
{\bf Supplementary Information} is available for this paper at
https://doi.org/10.1038/s41567-018-0228-3.\\[0.1cm]
\noindent{\bf Reprints and permission information} is available at 
www.nature.com/reprints.\\[0.1cm] 
\noindent{\bf Correspondence:} Correspondence and requests for materials
should be addressed to J.K. (email: kroha@th.physik.uni-bonn.de) 
or M. F. (email: manfred.fiebig@mat.ethz.ch).\\[0.1cm]
\noindent{\bf Publisher's note:} Springer Nature remains neutral with regard to 
jusrisdictional claims in published maps and institutional affiliations.

\vfill\eject

\noindent
{\bf Methods}

\noindent
{\bf Experiment.} The CeCu$_{6-x}$Au$_x$ samples used in this study were cut from single crystals, and faces perpendicular to the principal axes were polished using SiC. The specimens were mounted in a temperature-controlled Janis SVT-400 helium reservoir cryostat with Tsurupica windows. A Ti:sapphire laser (800~nm, 130~fs, 1~kHz, 2~mJ/pulse) generated single-cycle terahertz pulses of a few nJ by optical rectification in a 0.5~mm ZnTe(110) single crystal. terahertz radiation with spectral range between $0.1-3$~THz was incident onto the sample under 45$^{\circ}$ with the terahertz electric field parallel to its $b$ axis. Because of the low pulse energies and the high reflectivity \cite{Marabelli90} of CeCu$_{6-x}$Au$_x$, heating by our terahertz pulses is negligible.

Electrooptical sampling was performed using the fundamental Ti:sapphire laser pulse as probe pulse that was time-delayed by the sampling time $t$ with respect to the reflected terahertz wave. The terahertz and probe beams were collinearly focused onto a ZnTe(110) crystal. The terahertz-induced ellipticity of the probe light is measured using a quarter-wave plate, a Wollaston polarizer and a balanced photodiode. Output from the latter was analyzed with a lock-in amplifier. In order to increase the accessible time delay between terahertz and probe pulses, Fabry-Perot resonances from the faces of the 0.5~mm thin ZnTe(110) crystal were suppressed by extending the crystal with a 2.5~mm thick terahertz-inactive optically bonded ZnTe(100) crystal.

Prior to the experiment we identified all the terahertz reflexes generated by the optical components in order to avoid confusion with the true terahertz signal generated in the CeCu$_{6-x}$Au$_x$. A measurement of all these terahertz signals is summarized in Supplementary Information A. The earliest of these artefacts appeared at a delay of 10~ps, outside the range discussed in this work.

A set of data points from the electrooptical sampling represents the time trace of the terahertz field pulse between $t=-4.0$~ps and $t=+8.5$~ps. Each set was normalized by a factor such that its  integrated intensity (the sum of the square of all the normalized data points in the set) equals one, i.e.\ all traces are scaled to the same over-all reflected power. Such normalized data are shown in Fig.~\ref{thz}. They were used to derive the heavy-fermion quasiparticle weight and the Kondo temperature as described in the following. (Refer to Supplementary Information B for an in-detail technical description of this procedure.)

The power of the echo pulse was calculated by integrating the squared electric field of the normalized time traces over the interval from 3.5 to 8.5~ps where the echo pulse appears (see insets of Fig.~\ref{thz}). These raw data exhibit an offset, caused by noise, the residual tail of the undelayed main pulse, and the incoherent Kondo correlations \cite{Klein08} mentioned in the main text and discussed in detail in Supplementary Information B. In the range between 150 and 300~K, the Kondo weight of all our CeCu$_{6-x}$Au$_x$ samples is known to be zero so that only the temperature-independent offset is present. We subtract the integrated offset signal derived in this temperature range from our raw data to obtain $w_K^*$ as shown in Fig.~\ref{qcp}a. The Kondo temperature $T_K^*$ is derived from the relation $T_K^* = h/k_B \tau_K^*$. Here $\tau_K^*$ is taken as peak of the fitted envelope of the terahertz echo pulse according to equation~(\ref{eq:envelope}).

\noindent {\bf Theory.} The dynamical band structure of the terahertz-excited
heavy-fermion system, including the crystal-field satellite bands, was
calculated using a nonequilibrium generalization of the dynamical mean-field
theory (DMFT) \cite{Aoki14} for the multi-orbital Anderson lattice model with
six local orbitals, grouped in three Kramers doublets. These orbitals
represent the crystal-field-split $J=5/2$ ground-state multiplet of the Ce
$4f$ shell. An infinite onsite repulsion within the Ce $4f$ orbitals was
assumed, enforcing the overall single-electron occupancy of the Ce $4f$
shell. As the impurity solver of the DMFT, an auxiliary-particle
representation of the Ce $4f$ electron fields was employed within the
non-crossing approximation \cite{Ehm07,Klein08,Kroha98} (NCA), reaching down
to base temperatures (before terahertz excitation) of $T\approx
0.1~T_K^*$. The DMFT and NCA were generalized to the nonequilibrium case using the Keldysh technique \cite{Hettler98}. The efficient implementation of the NCA algorithm in Ref.~ \cite{Kroha98} was adapted for solving the DMFT with the multi-orbital NCA out of equilibrium.

\noindent{\bf Data availablity:} The data that support the plots within this
paper and other findings of this study are available from the corresponding
authors upon reasonable request.

%


\onecolumngrid

\

\newpage

\noindent
{\large{\bf\underline{Supplementary Information:}}}\\[0.3cm] 
{\large{\bf  
Time-resolved collapse and revival of the Kondo state 
near a quantum phase transition}}\\[0.2cm]

\noindent
C. Wetli$^{1}$, S. Pal$^{1}$, J. Kroha$^{2,3}$, K. Kliemt$^{4}$, 
C. Krellner$^{4}$, O. Stockert$^{5}$, 
H. v. L\"ohneysen$^{6}$, M. Fiebig$^{1}$\\[0.4cm]

\noindent
$^{1}$ Department of Materials, ETH Zurich, 8093 Zurich, Switzerland\\
$^{2}$ Physikalisches Institut and Bethe Center for Theoretical Physics,
Universit\"at Bonn, Nussallee 12, D-53115 Bonn, \phantom{$^2$}~Germany\\
$^{3}$ Center for Correlated Matter, Zhejiang University, 
Hangzhou, Zhejiang 310058, China\\
$^{4}$ Physikalisches Institut, Goethe-Universit\"{a}t Frankfurt, Max-von-Laue-Str.\ 1, 60438 Frankfurt, Germany\\
$^{5}$ Max Planck Institute for Chemical Physics of Solids, 01187 Dresden, Germany\\
$^{6}$ Institut f\"ur Festk\"orperphysik and Physikalisches Institut, Karlsruhe Institute of Technology, 76021 Karlsruhe,\\ \phantom{$^6$}~Germany

\vspace*{1cm}

\twocolumngrid

\setcounter{equation}{0}
\setcounter{figure}{0}
\setcounter{page}{1}
\renewcommand{\theequation}{S\arabic{equation}}
\renewcommand{\thefigure}{S\arabic{figure}}
\renewcommand{\thepage}{S\arabic{page}}

\begin{figure}[t]
\vspace*{0.3cm}
\centering
\includegraphics[width=0.75\linewidth]{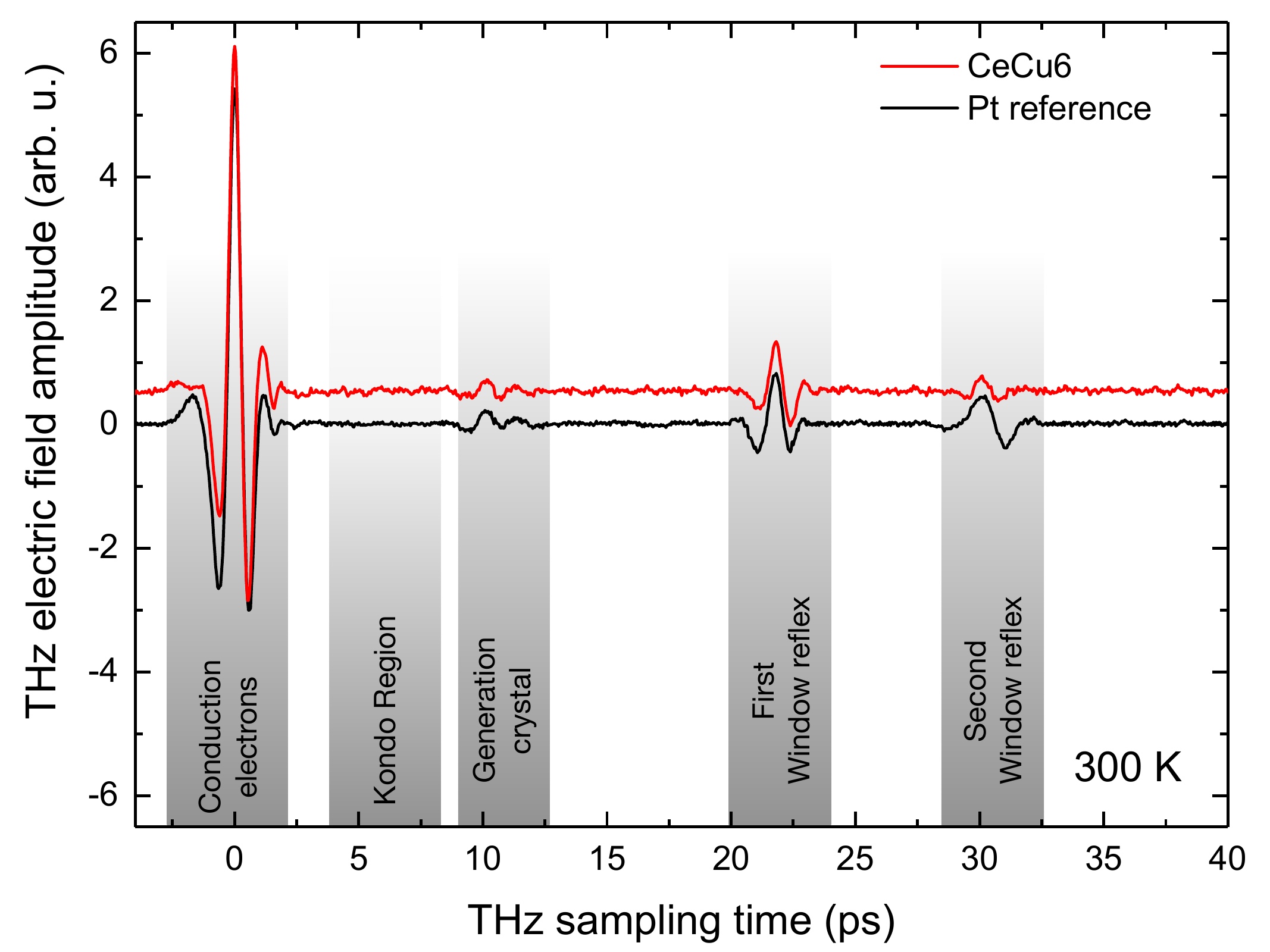}
\caption{\label{suppfig1} \textbf{Trivial reflections in the terahertz time traces.} Unnormalized terahertz electric field amplitudes obtained at room temperature from a reference Pt surface and the heavy-fermion CeCu$_6$ sample. Aside from the main instantaneous pulse at $t=0$, various reflex artefacts are observed whose origins are indicated in the plot. Note that no such artefact is present at a delay of 6~ps, the position of the Kondo echo pulse discussed in the main text. Furthermore, the reflex artifacts in the plot exhibit none of the complex temperature dependence of the echo pulse depicted in Fig.~4 of the main text.}
\end{figure}

\textbf{A) Identification of terahertz reflexes.} A time trace of 40~ps is taken to identify the signatures of optical components in our terahertz time-domain setup. These time traces are performed at room temperature with a CeCu$_6$ sample and a reference platinum mirror mounted within a cryostat and are shown in Fig.~\ref{suppfig1}. The instantaneous pulse at $t=0$ represents the light conduction electrons with relaxation times less than the duration of the pulse itself. In the case of the Pt reference, the next reflex that is observed is at 10~ps, which corresponds to the $d=0.5$~mm ZnTe (110) generation crystal and can be obtained using
\begin{equation}
  \Delta t = \frac{{2{n_{\rm THz}}d}}{c}
\end{equation}
where $c$ is the velocity of light and $n_{\rm THz}=3.2$ is the terahertz refractive index of ZnTe \cite{S_Schall99}. The other reflexes at 22~ps and 30~ps are due to the windows of the cryostat that are pairwise identical. The time trace from the Pt reference shows no signature around $t=6$~ps, the time where the characteristic Kondo echo pulse is observed in the CeCu$_{6-x}$Au$_{x}$ system. CeCu$_6$ shows the same trivial reflexes as the Pt mirror. By performing our experiments until $t=8.5$~ps, we ensure that no corrections for these trivial reflexes are necessary. The use of a 2.5~mm thick terahertz-inactive optically bonded ZnTe (100) crystal pushes the etalon effect from the detection crystal beyond the limit of $t=40$~ps shown in Fig.~\ref{suppfig1}.

The linearly polarized terahertz radiation used for excitation of the heavy-fermion systems covers the spectral range of about $0.4-12.5$~meV. Thus, excitations other than from the heavy-fermion band, like excitations of the Ce 4f single-particle bound states ($\sim 1$~eV) or spin-orbit transitions ($\gtrsim 100$~meV), can be neglected \cite{S_Klein08}.\\

\begin{figure}[t]
\centering
\includegraphics[width=0.95\linewidth]{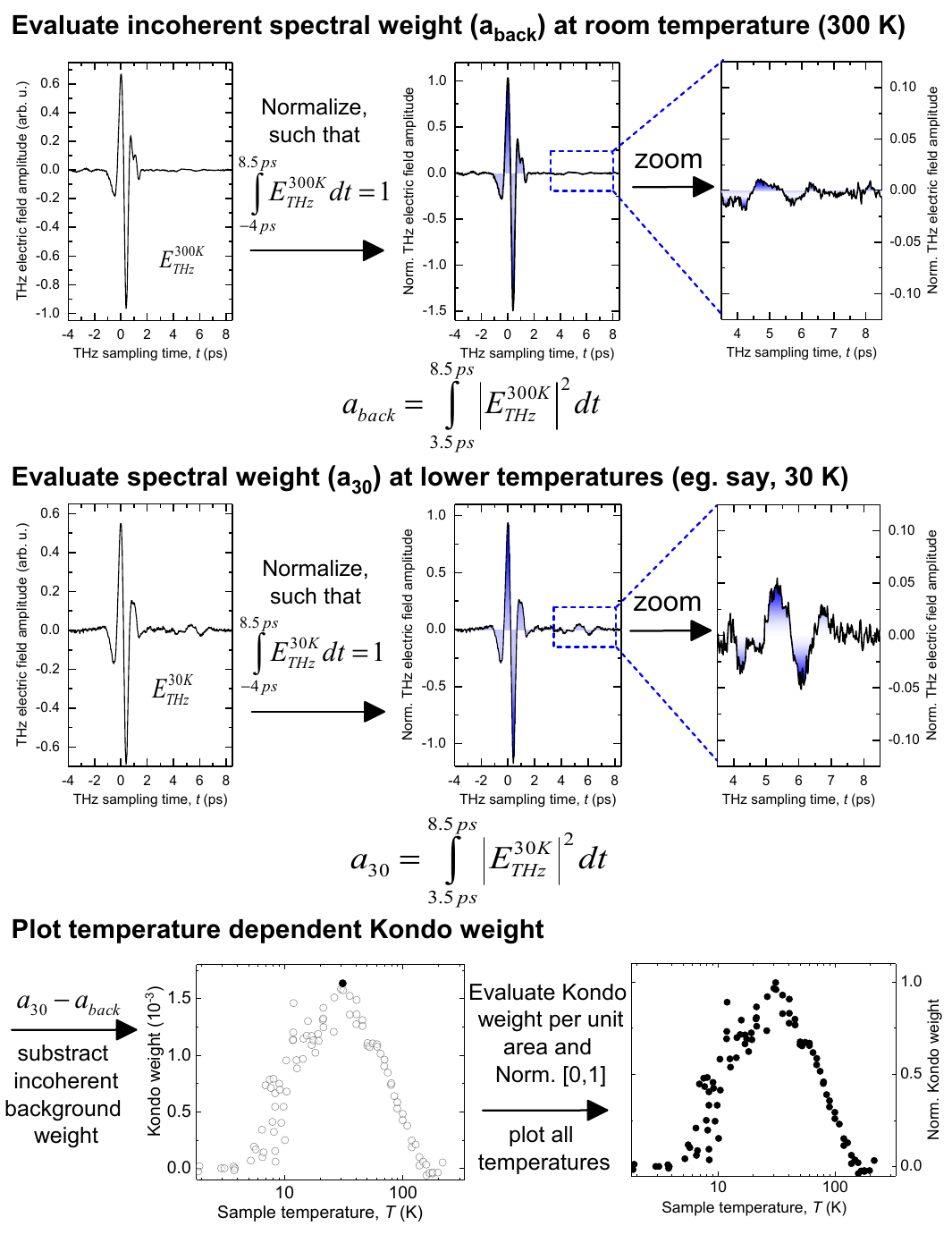}
\caption{\label{suppfig2} \textbf{Determination of the Kondo weight from the terahertz time-domain reflectometry data.} Sketch of the derivation procedure for an exemplary data point at $T=30$~K in the temperature-dependent Kondo weight. The procedure is discussed in detail in the text.}
\end{figure}

\textbf{B) Extracting the Kondo weight.} The procedure for extracting the coherent Kondo spectral weight $w_K^*$ is sketched as a step-by-step procedure in Fig.~\ref{suppfig2}. At first, all time traces of the terahertz electric field amplitude between $t=-4$~ps and $t=8.5$~ps are normalized by a factor in order to scale all of them to an over-all identical total reflected power. The normalization is done such that the integrated intensity (i.e., the sum of the square of all the normalized and equidistant data points in the set) equals one.

The normalized data are then used to derive the heavy-fermion quasiparticle weight as follows: The power of the echo pulse, represented by $a_{30}$ in Fig.~\ref{suppfig2} for an exemplary data point at 30~K, is calculated by integrating the squared electric field of the normalized time traces over the interval from 3.5 to 8.5~ps, where the echo pulse appears. The choice of this time window is somewhat aided by the envelope function that is the solution of the non-linear rate equation model, expressed by equation~(2) in the main text. Technically, 8.5~ps is the upper limit since the etalon effect from the generation crystal follows next, starting from 8.9~ps. We observed that shifting the lower limit of the integration window merely introduces more noise, yet without adding further spectral weight.

In all samples the integrated power of the echo pulse exhibits a temperature-independent offset. This offset is caused by experimental noise, the residual tail of the instantaneous main reflex and, in particular, to incoherent spectral-weight contributions to the Kondo physics, to be discussed below. Above 150~K, where the Kondo coherence is known to be lost, this offset is the only contribution to the integration of the squared time traces. Hence, for each sample, we determine the offset value (represented by $a_{\rm back}$ in Fig.~\ref{suppfig2}) by calculating the power of the echo pulse as described before, yet at 300~K. The offset value is then subtracted from the raw data to obtain the coherent quasiparticle weight at a specific temperature, in our exemplary case $a_{30}-a_{\rm back}$ at 30~K in Fig.~\ref{suppfig2}. Finally, the spectral weight curves of the CeCu$_{6-x}$Au$_{x}$ samples in Fig.~4a are normalized to the sample area on which the terahertz data were taken. This last normalization allows for easier comparison and eliminates variations resulting from the fact that the sample areas probed on the different CeCu$_{6-x}$Au$_{x}$ samples are markedly different.

\begin{figure}[t]
\centering
\includegraphics[width=0.75\linewidth]{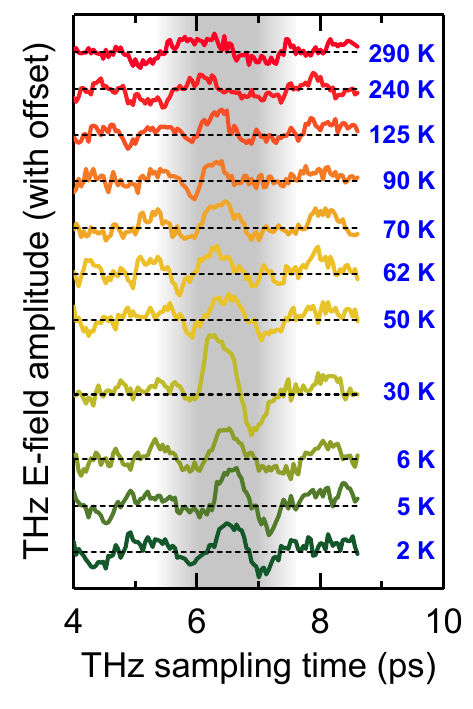}
\caption{\label{suppfig3} \textbf{Temperature-dependent terahertz time traces for CeCu$_6$.} At $T=30$~K the oscillatory terahertz echo pulse emitted after about 6~ps is clearly visible (area shaded in grey). However, experimental noise, the residual tail of the instantaneous main reflex and incoherent Kondo correlations \cite{S_Klein08} contribute to this delayed pulse, too, and need to be subtracted in the calculation of the coherent Kondo weight. Above about 150~K, the coherent Kondo weight is known to be zero so that only the temperature-independent offset remains. Data taken in this temperature range can therefore be used to determine the offset value. The dashed lines mark the vertically displaced zero-intensity lines of the scans.}
\end{figure}

As mentioned above, in heavy-fermion compounds incoherent spectral weight is still present outside of the heavy-Fermi-liquid regime, e.g.\ in CeCu$_{6-x}$Au$_{x}$ at high temperatures, as a feature whose spectral width exhibits the Kondo scale. This incoherent background was previously observed in photoemission spectroscopy experiments \cite{S_Klein08}. In our experiments it appears as a time-delayed enhancement of the noise in the background signal, visible e.g.\ in Fig.~\ref{suppfig3}. Thus, analyzing these background features allows us to extract the Kondo scale $T_K^*=h/k_B\tau_K^*$ even in situations where the coherent Kondo weight has vanished.\\

\textbf{C) Non-linear rate equation model.} The electric-field amplitude $\overline{E}(t)$ of the emitted terahertz pulse is proportional to the temporal change of the density $\rho(t)$ of photoelectrons in the conduction band,
\begin{equation}\label{eq:rate}
  \overline{E}(t) \sim \frac{\mathrm{d}\rho(t)}{\mathrm{d}t} = -\Gamma_{c\to hf}\,\rho(t)\,,
\end{equation}
where $\Gamma_{c\to hf}$ is the transition rate of light conduction ($c$) electrons to the heavy-fermion ($hf$) band. Using Fermi's golden rule, $\Gamma_{c\to hf}$ is expressed in terms of the dipole transition matrix element $\langle f_{hf}|d|i_c\rangle$ and of the spectral density in the final states of the transition. This spectral density corresponds to the weight $\tilde w_K^*$ of the heavy-fermion band, so that we get
\begin{equation}\label{eq:golden_rule}
  \Gamma_{c\to hf} = \frac{2\pi}{\hbar}\,|\langle
  f_{hf}|d|i_c\rangle|^2\,\tilde w_K^*.
\end{equation}
Resulting from a many-body correlation effect, $\tilde w_K^*$ is, however, not a static quantity, but proportional to the number of Kondo singlets forming this band, that is, to the density of electrons residing in the heavy-fermion band at time $t$,
\begin{equation}\label{eq:HFweight}
  \tilde w_K^*(t) = w_K^*\,\frac{\rho_0-\rho(t)}{\rho_0},
\end{equation}
where $w_K^*$ and $\rho_0$ are the equilibrium values of the Kondo spectral weight and of the electron density in the heavy-fermion band, respectively. We describe the Kondo resonance in a standard way \cite{S_Hewson93} as a Lorentzian of width $\sim k_BT_K^*$ and height $\sim 1/(\pi\gamma)$ with $N_c$ as the conduction electron density of states in the volume $V$. Further, $\gamma=\pi V^2N_c$ is the effective hybridization amplitude of the Anderson impurity model \cite{S_Hewson93}. It may be estimated from the dipole matrix element as $\gamma\approx \pi |\langle f_{hf}|d|i_c\rangle|^2 N_c$. Putting this together, we derive an estimate for the equilibrium heavy-fermion spectral weight as
\begin{eqnarray}\label{eq:HFweight0}
  w_{K}^* = \frac{k_BT_K^*}{\gamma}\,N_c \,.
\end{eqnarray}
Combining equations~(\ref{eq:rate})--(\ref{eq:HFweight0}) and using $k_BT_K^*=2\pi\hbar/\tau_K^*$ thus yields a non-linear rate equation for the normalized density of photo-excited electrons, $n(t)=\rho(t)/\rho_0$,
\begin{eqnarray}\label{S_eq:rate_eqn}
  \frac{\mathrm{d}n}{\mathrm{d}t}=-\frac{4\pi}{\tau_K^*}\,(1-n)\,n.
\end{eqnarray}
We see that the non-linearity arises from from the fact that the heavy-fermion spectral weight is dynamical: $\tilde w_K^*\equiv {\tilde w_K^*}(t)$. We thus have a convolution of the recreation of the ground-state spectral density and the repopulation of this recreating spectral density. The overall small value of $w_K^*\sim T_K^*$ in comparison to the light conduction electron spectral weight $N_c$ results in an effective relaxation constant $4\pi/\tau_K^*$, strongly reduced with respect to the bare dipole matrix element. The solution of equation~(\ref{S_eq:rate_eqn}) reads
\begin{equation}\label{eq:n_t}
  n(t) = \frac{1}{2}\left[ 1- \tanh(2\pi (t-t_0)/\tau_K^*) \right]\,.
\end{equation}
For the electric-field envelope of the terahertz reflex this yields
\begin{equation}\label{S_eq:envelope}
  \overline{E}(t) =\frac{\overline{E}_0}{\cosh^2[2\pi (t/\tau_K-1)]}.
\end{equation}

\vfill
\eject

\onecolumngrid

\twocolumngrid

\vfill

\end{document}